\documentclass[12pt,letter]{article}
\pdfoutput=1
\usepackage{graphicx, epsfig, color,cite}
\usepackage{amsmath}
\usepackage{amssymb}
\usepackage{float}
\usepackage{subfig}
\usepackage{hyperref}

\def\to{\rightarrow}

\def\bi{\begin{itemize}}
\def\ei{\end{itemize}}

\def\ta{\tilde a}
\def\tchi{\tilde\chi}

\def\th{\tilde h}
\def\tst{\tilde t}

\def\tg{\tilde g}

\def\alt{\lesssim}
\def\agt{\gtrsim}
\def\be{\begin{equation}}  
\def\ee{\end{equation}}  
\def\bea{\begin{eqnarray}}  
\def\eea{\end{eqnarray}}

\begin{document}
\begin{titlepage}
\begin{flushright}
OU-HEP-260702
\end{flushright}

\vspace{0.5cm}
\begin{center}
  {\Large \bf Can blind spots save neutralino dark matter\\
  in natural supersymmetry models?}\\
\vspace{1.2cm} \renewcommand{\thefootnote}{\fnsymbol{footnote}}
{\large Howard Baer$^{1}$\footnote[1]{Email: baer@ou.edu },
Vernon Barger$^2$\footnote[2]{Email: barger@pheno.wisc.edu}
and Dibyashree Sengupta$^{3,4}$\footnote[4]{Email: sengupta.dibyashree@ucy.ac.cy}
}\\ 
\vspace{1.2cm} \renewcommand{\thefootnote}{\arabic{footnote}}
{\it 
$^1$Homer L. Dodge Department of Physics and Astronomy,
University of Oklahoma, Norman, OK 73019, USA \\[3pt]
}
{\it 
$^2$Department of Physics,
University of Wisconsin, Madison, WI 53706 USA \\[3pt]
}
{\it
$^3$ INFN, Sezione di Roma, c/o Dip. di Fisica, Sapienza Università di Roma, Piazzale Aldo Moro 2, I-00185 Rome, Italy} \\[3pt]
{\it
$^4$ Department of Physics, University of Cyprus, P.O. Box 20537, 1678 Nicosia, Cyprus} \\[3pt]
\end{center}

\vspace{0.5cm}
\begin{abstract}
\noindent
Natural supersymmetry (SUSY) models remain viable even in the face of
LHC Run 2 sparticle search limits.
However, the LZ experiment has placed strong limits
on light higgsino dark matter even when the higgsinos carry only their
thermally-produced abundance, with the bulk of the dark matter composed of
axions.
One way out is the possibility of WIMP direct detection blind spots where
cancellations in direct detection (DD) couplings lead to tiny DD rates.
We examine natural SUSY models with $\mu <0$ and $\mu >0$
but find that the surviving blind spots
all lie in the unnatural region where the superpotential $|\mu |$
parameter is much greater than the weak scale gaugino masses; the few
natural candidates are excluded by LHC soft-dilepton searches and by the
measured Higgs mass.
Within NUHM2/NUHM3-type gravity-mediated models with
positive gaugino masses and assuming a thermally produced neutralino
fractional abundance, direct-detection blind spots do not rescue stable
light higgsino dark matter in the electroweak-natural region.
Thus, within this framework, stable light higgsino dark matter
is disfavored, although special circumstances like large entropy dilution
of all relics is still possible.
This points to SUSY models with {\it unstable} light higgsinos as
perhaps the preferred alternative.

\end{abstract}
\end{titlepage}

\section{Introduction}
\label{sec:intro}

Weak scale supersymmetry (SUSY) was proposed in the early 1980s as
an extension of the Standard Model (SM) 
which is free of the destabilizing quadratic divergences that plague the
Higgs mass\cite{Witten:1981nf,Kaul:1981wp,Dimopoulos:1981zb} (for textbook treatment, see {\it e.g.} Ref. \cite{Baer:2006rs}).
From a string theory perspective, 
the $N=1$ spacetime SUSY emerges from Calabi-Yau
string compactifications on Ricci-flat manifolds of special holonomy\cite{Candelas:1985en}.
Non-SUSY compactifications have been argued to lead to instabilities
such as Witten's bubble of nothing\cite{Acharya:2019mcu}.

\subsubsection{The naturalness issue and the $\Delta_{EW}$ measure}

In pre-LHC times, naturalness considerations seemed to imply that
superpartner masses within the context of the Minimal Supersymmetric
Standard Model (MSSM) should lie at or around the weak scale\cite{Feng:2013pwa}.
When supersymmetric matter failed to turn up in early LHC searches,
greater scrutiny was placed on the naturalness question.
\bi
\item The log-derivative measure of finetuning\cite{Ellis:1986yg,Barbieri:1987fn}
$\Delta_{p_i}\equiv max_i|\partial\log m_Z^2/\partial\log p_i |$ asked for small
variations in $m_Z^2$ with respect to variation in high scale soft SUSY breaking
parameters.
However, neglecting correlations between the soft parameters
(which in popular models such as mSUGRA/CMSSM are input as independent
parameters of ignorance whereas in explicit constructs they all emerge as
computable numbers in terms of the gravitino mass $m_{3/2}$) leads to
finetuning overestimates\cite{Baer:2013gva,Mustafayev:2014lqa} in the range of
$10-10^3$\cite{Baer:2023cvi}.
\item Likewise, the measure $\Delta_{HS}\equiv \delta m_{H_u}^2/m_h^2$
  neglects the fact that the logarithmic evolution of $m_{H_u}^2$ is
  dependent on the value of the soft term $m_{H_u}^2$ itself, and furthermore
  its evolution, for phenomenological viability, must be large enough
  to result in a (radiative-driven) breakdown of  electroweak symmetry\cite{Ibanez:1982fr,Alvarez-Gaume:1983drc}.
  Use of $\Delta_{HS}$ also results in finetuning overestimates by
  factors of $\sim 10-10^3$\cite{Baer:2023cvi}.
\ei

A more conservative, model-independent measure of finetuning $\Delta_{EW}$
was proposed in Ref. \cite{Baer:2012up,Baer:2012cf}.
Minimization of the MSSM scalar potential allows one to relate the
measured value of $m_Z$ to the weak scale MSSM Lagrangian parameters:
\be
m_Z^2/2=\frac{m_{H_d}^2+\Sigma_d^d-(m_{H_u}^2+\Sigma_u^u)\tan^2\beta}{\tan^2\beta -1}
  -\mu^2\simeq -m_{H_u}^2-\mu^2-\Sigma_u^u(\tst_{1,2})
  \label{eq:mzs}
\ee
where $m_{H_u}^2$ and $m_{H_d}^2$ are soft SUSY breaking Higgs masses,
$\mu$ is the (SUSY-conserving) Higgs/higgsino superpotential mass term\footnote{Twenty solutions to the SUSY $\mu$ problem are reviewed in Ref. \cite{Bae:2019dgg}.},
$\tan\beta \equiv v_u/v_d$ is the ratio of Higgs field vacuum expectation values
and the $\Sigma_u^u$ and $\Sigma_d^d$ terms contain an assortment of radiative
corrections, the most important of which are usually the
$\Sigma_u^u(\tst_{1,2})$\cite{Baer:2012cf}.
The numerical measure $\Delta_{EW}$ is defined by
\be
\Delta_{EW}\equiv max_i| i^{th}\ term\ on\ RHS\ of\ Eq.~\ref{eq:mzs}|/(m_Z^2/2) .
\ee
A value of $\Delta_{EW}\alt 30$ requires each independent contribution
to $m_Z$ to be within a factor of four of $m_Z$ itself.
This upper limit on $\Delta_{EW}$ coincides with the so-called anthropic limit
of the weak scale as derived by Agrawal {\it et al.} (the ABDS window)\cite{Agrawal:1997gf}. SUSY models with a large value of $\Delta_{EW}$ suffer from a little hierarchy problem
(LHP) whilst natural SUSY models with low $\Delta_{EW}$ do not.

Several implications of $\Delta_{EW}\alt 30$ are the following.
\bi
\item Since the $\mu$ parameter is directly bounded, then the higgsino-like
  electroweakinos of the MSSM should have mass values $m_{\tchi_{1,2}^0}$ and
  $m_{\tchi_1^\pm}$ $\sim 100-350$ GeV. Thus, unlike most $20^{th}$ century SUSY
  constructs, the higgsinos are the lightest SUSY particles.
\item The weak scale value of $m_{H_u}^2$ is radiatively-driven to small
  negative values $\sim -(100-350\ {\rm GeV})^2$. This usually requires
  non-universal Higgs masses, as is generically expected in actual
  supergravity constructs\cite{Soni:1983rm,Kaplunovsky:1993rd,Brignole:1993dj}.
\item The (usually) dominant radiative corrections $\Sigma_u^u (\tst_{1,2})$
  are minimized by large, negative trilinear soft terms $A_0$\cite{Baer:2012up}.
  Large trilinears are also generic in SUGRA constructs.
  They also help to uplift the light Higgs mass $m_h\to 125$ GeV\cite{Carena:2002es}.
  Under $\Delta_{EW}\alt 30$, top squarks may range up to $m_{\tst_1}\alt 3$ TeV
  and $m_{\tst_2}\alt 8$ TeV\cite{Baer:2015rja}.
\item The gluinos contribute to the $\Sigma_{u,d}^{u,d}$ terms at two-loop
  level and thus can range up to 6-9 TeV (model-dependent\cite{Baer:2015rja,Baer:2018hpb}).
\item First/second generation squark and slepton mass contributions
  to the weak scale are Yukawa-coupling suppressed.
  Thus, they can range up to 10-40 TeV.
  These latter values can lead to a mixed decoupling/quasi-degeneracy solution
  to the SUSY flavor and CP problems\cite{Dine:1993np,Cohen:1996vb,Baer:2019zfl}
  which would otherwise afflict gravity mediation models.
\ei

\subsection{WIMP dark matter in natural SUSY}

From the above considerations of naturalness, one expects the lightest
SUSY particle (LSP) to be a higgsino-like lightest neutralino (as opposed to pre-LHC
theoretical prejudice which expected a bino-like LSP).
Higgsino-like LSPs in the natural SUSY mass range of $m_{\tchi}\sim 100-350$ GeV
are well-known to be thermally underproduced as a dark matter candidate,
with $\Omega_{\th}h^2\sim 0.01$, typically a factor 10 below the measured abundance.
A higgsino-like LSP saturates the measured DM abundance for $m_{\th}\sim 1.1$ TeV,
well above the naturalness limit.

Of course, if one insists on naturalness in the QCD sector, then one is led to the
axion solution to the strong CP problem.
In natural SUSY models that include the QCD axion,  then one expects the presence of
SUSY partners the spin-1/2 axino $\ta$ and spin-0 saxion $s$,
where generic supergravity expectations are for $m_{\ta}\sim m_s\sim m_{3/2}$.
In $R$-parity conserving (RPC) models, then dark matter can consist of a
a neutralino/axion admixture\cite{Baer:2011hx}. In such a case, then WIMPs may comprise
only a fraction $\xi$ of the dark matter, where $\xi\equiv \Omega_{\tchi}h^2/0.12$.
For natural SUSY models with thermally-produced light higgsinos,
one then expects $\xi\sim 0.1$, and direct detection bounds would need to be rescaled
by a factor $\xi$ reflecting the diminished local abundance of WIMP dark matter.

In RPC SUSY axion models, one usually doesn't expect the light higgsinos to obtain
their calculated thermal abundance. The reason is that thermal production of axinos
and thermal- and coherent-oscillation production of saxions, followed by
axino/saxion decay to LSPs in the early universe, can augment the thermal abundance.
Also, if saxions mainly decay to SM particles instead of SUSY particles, then
entropy dumping from saxion decay can dilute any relics present at the time of
saxion decay. Thus, it is possible to either increase or decrease the neutralino
abundance from  its thermally-produced expectation in SUSY axion models.
In addition, the presence of stringy-motivated light moduli fields in the early
universe may also augment or diminish the expected WIMP abundance.
In a previous work, it was suggested that light natural higgsino-like WIMP-only dark
matter is already excluded by a combination of direct
and indirect (IDD) WIMP dark matter detection experiments\cite{Baer:2018rhs}.
Recently, the LZ experiment\cite{LZ:2024zvo} has published strong new WIMP DD limits
which for instance require $\xi\sigma^{SI}(\tchi p )\alt 5\times 10^{-12}$ pb
for $m_{\tchi}\sim 200$ GeV.
These limits nearly exclude light, higgsino-like WIMP dark matter even in the case
where the WIMPs make up only their thermally-produced relic abundance\cite{Baer:2016ucr}.

The question then arises:
is natural SUSY excluded, or nearly excluded, by the new LZ
DD bounds, especially when combined with the LHC measured value of the
Higgs boson mass and sparticle mass limits?
An answer of {\it no} was found recently in Ref's
\cite{Baer:2025oid} and \cite{Baer:2025srs}.
In these works, if the SUSY $\mu$ term is forbidden via the introduction
of discrete $R$-symmetries, and then is regenerated via the
Kim-Nilles mechanism\cite{Kim:1983dt} via the non-renormalizable operator
\be
W\ni X^pY^q H_uH_d/m_P^{p+q-1}\ \ \ \ \ (Kim-Nilles)
\ee
when the PQ-charged fields $X$ and $Y$ receive vevs of order
$\sim f_a\sim m_{hidden}\sim 10^{11}$ GeV,
then renormalizable $R$-parity violating operators are also forbidden.
However, non-renormalizable RPV operators of the form
\be
W\ni X^rY^s QQQ/m_P^{r+s}
\ee
can now be allowed and will lead to RPV-operators with suppressed coefficients of order
$\lambda_{RPV}\sim (f_a/m_P)^n$. One possibility is the case where $n=1$, so
$\lambda_{RPV}\sim 10^{-7}$. In this case, the WIMPs can be produced in the early universe, but
will all decay before the onset of Big-bang nucleosynthesis (BBN), yielding a SUSY
model with {\it all axion} dark matter. Another possibility is that the SUGRA model features
a conspiracy of parameters such that the axino is light and is itself the LSP\cite{Chun:2020fad}.
In this case,
the lightest higgsino-like neutralino decays again with a delayed decay via
$\tchi\to h\ta$ or $Z\ta$ and one has mixed axion/axino dark matter\cite{Baer:2026wre}.

A third means of getting around the strong LZ constraints is examined in this paper,
and that is via the possibility of {\it blind spots} in the DD scattering
rates\cite{Baer:1997ai,Ellis:2000ds,Baer:2003jb,Baer:2006te,Baer:2008ih,Cheung:2012qy,Huang:2014xua,Han:2016qtc,Bae:2022dgj,Arcadi:2025sxc}.
Blind spots have been examined for over 25 years, and can arise via a combination of
MSSM model parameters which lead to cancellations in the DD couplings of neutralinos to either
$h$ or $Z$. For our case of natural SUSY with decoupled first/second generations scalars,
then direct detection of WIMPs via SI interactions takes place primarily via light Higgs exchange,
with a coupling (in the notation of Ref. \cite{Baer:2006rs})
\be
{\cal L}\ni X_{ij}^h\bar{\tchi}_i^0 (-i\gamma_5)^{\theta_i+\theta_j}\tchi_j^0 h
\ee
where
\be
X_{ij}^h =-\frac{1}{2}(-1)^{\theta_i+\theta_j}\left(v_2^{(i)}\sin\alpha-v_1^{(i)}\cos\alpha\right)\left(gv_3^{(j)}-g^\prime v_4^{(j)}\right)
\ee
where $v_1^{(i)}$ and $v_2^{(i)}$ are higgsino components of the $i$th neutralino,
$v_3^{(j)}$ and $v_4^{(j)}$ are the wino and bino components of the $j$th neutralino,
$\alpha$ is the Higgs mixing angle and the $\theta_i$ are 0 or 1 depending on the sign of
the $i$th neutralino mass eigenstate (whether it needs a chiral rotation to assume a
canonical mass term).
This coupling is typically maximal for a highly mixed (well-tempered\cite{Arkani-Hamed:2006wnf,Baer:2006te})
neutralino (now excluded\cite{Baer:2016ucr}), and goes to zero for either a pure higgsino or gaugino (neither of which is obtained in realistic
SUSY models).
But depending on signs and magnitudes of the neutralino mixing components, the
$X_{ij}^h$ coupling can become highly suppressed leading to DD blind spots.

Cheung {\it et al.} find that the SI blind spot condition factorizes\cite{Cheung:2012qy,Han:2016qtc} as
\be
\left(m_{\tchi}+\mu \sin 2\beta\right)\left(m_{\tchi}-\frac{1}{2}\left[M_1+M_2+(M_1-M_2)\cos 2\theta_W\right]\right)=0 ,
\ee
which is satisfied approximately in four cases:
\begin{enumerate}
\item bino LSP with $M_1+\mu \sin 2\beta =0$ and $sign (M_1/\mu )=-1$,
  \item wino LSP with $M_2+\mu \sin 2\beta =0$ and $sign (M_2/\mu )=-1$,
    \item higgsino LSP with $\tan\beta \sim 1$ and $sign (M_{1,2}/\mu )=-1$ and
      \item the bino/wino degenerate case $m_{\tchi}=M_1=M_2$ arising from the
        second factor (independent of the sign of $\mu$).
\end{enumerate}
For SUGRA models with positive gaugino masses, this points to examination of
cases with $\mu <0$, which tended not to be emphasized when the
$(g-2)_\mu$ anomaly was present
($\mu <0$ pushed $(g-2)_\mu$ in the wrong direction compared to the
old measured anomaly\cite{Moroi:1995yh,Baer:2001kn}).
Since the $(g-2)_\mu$ anomaly has seemingly disappeared\cite{Aliberti:2025beg},
there is now little or no prejudice against $\mu <0$.
Blind spots for intermediate sub-TeV values of $m_A$ have been considered in
Ref. \cite{Huang:2014xua}. Current limits on $m_A$ from LHC searches for
$H,A\to\tau^+\tau^-$ decay imply $m_A\agt 0.8$ TeV (for $\tan\beta\sim 10$) and
$m_A\agt 2$ TeV (for $\tan\beta\sim 50$)\cite{Baer:2022qqr}. Thus, here we restrict ourselves
to heavier $m_A$ values beyond the sub-TeV range in accord with LHC heavy Higgs
search limits.

In this paper, we examine whether DD blind spots which appear
for SUGRA models with gaugino masses $M_{1,2}>0$ and with $\mu <0$ can save
natural higgsino-like WIMP dark matter from the strong LZ DD limits.
In Sec. \ref{sec:comp}, we give a brief description of our calculational
methods.
In Sec. \ref{sec:nuhm}, we show natural SUSY regions and blind spot
regions of the $\mu$ vs. $m_{1/2}$ parameter space of the
two-and-three-extra-parameter non-universal Higgs model (NUHM2,3).
In Sec. \ref{sec:gen}, we present results from a more general scan over
parameter space. In Sec. \ref{sec:conclude}, we give a summary along with our conclusion: no, blind spots cannot save $R$-parity conserving natural
SUSY from the strong LZ DD bounds (except for a few exceptional cases).
This points to models with a newer paradigm of all axion or mixed axion/axino
dark matter as perhaps a favored scenario for SUSY dark matter.

\section{Computational details}
\label{sec:comp}

For our SUSY spectrum generation, we use Isajet 7.92\cite{Paige:2003mg}
which features Isasugra\cite{Baer:1994nc} for the spectrum calculation.
Isasugra includes one-loop sparticle mass corrections following
Pierce {\it et al.}\cite{Pierce:1996zz}.
For neutralinos, the complete one-loop mass corrections are included,
while for v7.92, we adopt the loop-corrected values of wino mass $M_2$
and higgsino mass $\mu$ into the chargino mass matrix to gain a consistent
set of loop-corrected electroweakino masses.
This prescription is advocated by Pierce {\it et al.} to give chargino masses to
$\sim 1\%$ precision.

For neutralino relic density, we use the Isatools code IsaReD\cite{Baer:2002fv}
which calculates all neutralino annihilation and coannihilation processes
using CalcHEP\cite{Pukhov:2004ca} and then  integrates to gain the
thermally-produced neutralino relic density $\Omega_{\tchi}^{TP}h^2$.
For neutralino DD rates, we use the code
IsaReS (Isajet Relic Scatter) as described in Ref. \cite{Baer:2003jb}.

\section{WIMP direct detection blind spots in the NUHMi models}
\label{sec:nuhm}

\subsection{Results in NUHM2 model}

To display DD blind spots in natural SUSY parameter space, we first adopt the
paradigm NUHM2 model\cite{Ellis:2002iu,Baer:2005bu} included in Isajet
with the six parameters
\be
m_0(1,2,3),\ m_{1/2},\ A_0,\ \tan\beta,\ \mu\ and\ m_A\ \ \ \ (NUHM2)
\ee
which unlike many other SUSY models allows for low $\Delta_{EW}\alt 30$ even
while respecting $m_h\sim 125$ GeV and LHC sparticle mass limits.
The non-universal Higgs masses $m_{H_u}^2$ and $m_{H_d}^2$ ($\ne m_0(1,2,3)$)
have been traded for the more convenient weak scale parameters
$\mu$ and $m_A$. The NUHM4 model (with $m_0(1)\ne m_0(2)\ne m_0(3)$)
is in accord with general expectations of SUGRA models where
non-universality is expected (unlike the CMSSM model), and where
$m_0(1,2)\gg m_0(3)$ offers a mixed quasi-degeneracy/decoupling solution
to the SUSY flavor and CP problems\cite{Baer:2019zfl}.
Here we take the various generations of matter scalars to be degenerate
as in NUHM2 as a simplification.

\begin{figure}[htb!]
\centering
    {\includegraphics[height=0.5\textheight]{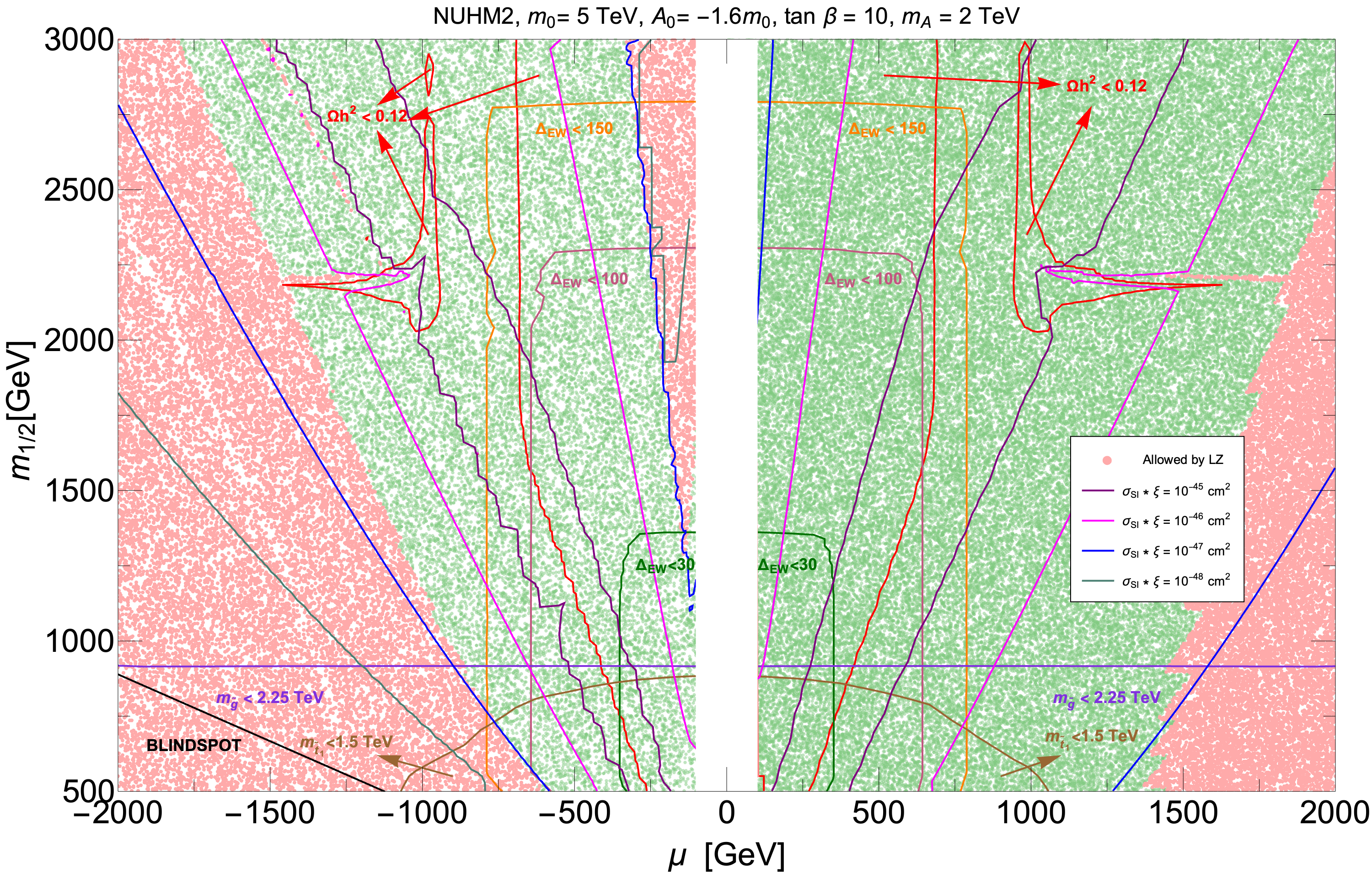}}
    \caption{The $\mu $ vs. $m_{1/2}$ parameter space of the NUHM2 model
      for $m_0=5$ TeV, $A_0=-1.6 m_0$, $\tan\beta =10$ and $m_A=2$ TeV.
      We show contours of $m_{\tg}$, $m_{\tst_1}$, $\Delta_{EW}$,
      $\Omega_{\tchi}^{TP}h^2$ and $\xi \sigma^{SI}(p\tchi )$.
      \label{fig:mu_v_mhf}}
\end{figure}

Our first results are plotted in the $\mu$ vs. $m_{1/2}$ plane of the NUHM2
model in Fig. \ref{fig:mu_v_mhf}. The assumption of gaugino masses
unified to $m_{1/2}$ is actually more generally motivated than just by
grand unified theories (GUTs):
if gaugino masses arise in SUGRA from a gauge kinetic function
which is a simple linear function of soft breaking fields,
then universal gaugino masses should ensue.
Also for this plot, we set $A_0=-1.6 m_0$
(to gain $m_h\simeq 125$ GeV over the bulk of the plot), $\tan\beta =10$,
$m_0=5$ TeV and $m_A=2$ TeV.

In Fig. \ref{fig:mu_v_mhf}, the region below the purple contour
labeled $m_{\tg}<2.25$ TeV is excluded by ATLAS/CMS searches for
gluino pair production. The region below the brown contour
is excluded by ATLAS/CMS searches for top-squark pair production
(within the context of simplified models). In the Figure, we also show
regions of $\Delta_{EW}<30,\ 100$ and $150$, with $\Delta_{EW}<30$ as the
suggested natural SUSY boundary where there is no LHP (low $|\mu |\alt 350$ GeV
and low $m_{1/2}\alt 1300$ GeV region). Inside the red contours
is where $\Omega_{\tchi}^{TP}h^2<0.12$. This includes the regions of low
$|\mu |$ where we obtain higgsino-like WIMPs, but also regions where
there is bino annihilation through the $A,H$ resonance (around $m_{1/2}\sim 2250$ GeV) and higgsino pair annihilation (around $|\mu|\sim 1000$ GeV).
The remaining regions would have too great a TP WIMP relic density, and
would need non-thermal processes like entropy dilution to gain accord with
the measured dark matter density.

Under RPC-SUSY with relic neutralino dark matter, the green points
would be excluded by the new LZ DD limits while the salmon (LZ-allowed)
points would be allowed.
There is a band of allowed points at small negative $\mu$ which
barely touches the natural $\Delta_{EW}\alt 30$ region.
This is one of the blind spots where DD rates are very low because the
$\tchi_1^0$ is nearly pure higgsino.
These so-called ``allowed'' points are likely excluded by
ATLAS\cite{ATLAS:2019lng} and CMS\cite{CMS:2021edw} limits on higgsino pair
production followed by decay to soft, opposite-sign isolated dileptons
(for which there are at present $\sim 2\sigma$ excesses from LHC Run 2 data).
There is also a blind spot at large $|\mu |$ and smaller $m_{1/2}$ where one
gains highly pure bino-like LSPs.
This is even more pronounced for $\mu <0$ where
$\xi\sigma^{SI}(\tchi ,p)$ drops below $10^{-48}$ cm$^2$:
one of the Cheung {\it et al.} blind spots.
The Cheung {\it et al.} blind spot for bino-like dark matter and
$\tan\beta \sim 10$ occurs at $\mu\sim -2.25 m_{1/2}$ and is indicated by a
dark line labeled BLINDSPOT.

Notice these large $|\mu |$ blind spots are well outside the naturalness
region with $\Delta_{EW}\alt 30$.
From this plot, and for this particular configuration of parameters, we would
conclude that natural SUSY with thermally-produced stable higgsino-like WIMPs
for the most part cannot escape the LZ bounds except for the case where
non-thermal processes like entropy dilution reduces $\xi$ to below its
TP value.

\subsection{Results in NUHM3 model}
\label{ssec:nuhm3}

\begin{figure}[htb!]
\centering
    {\includegraphics[height=0.5\textheight]{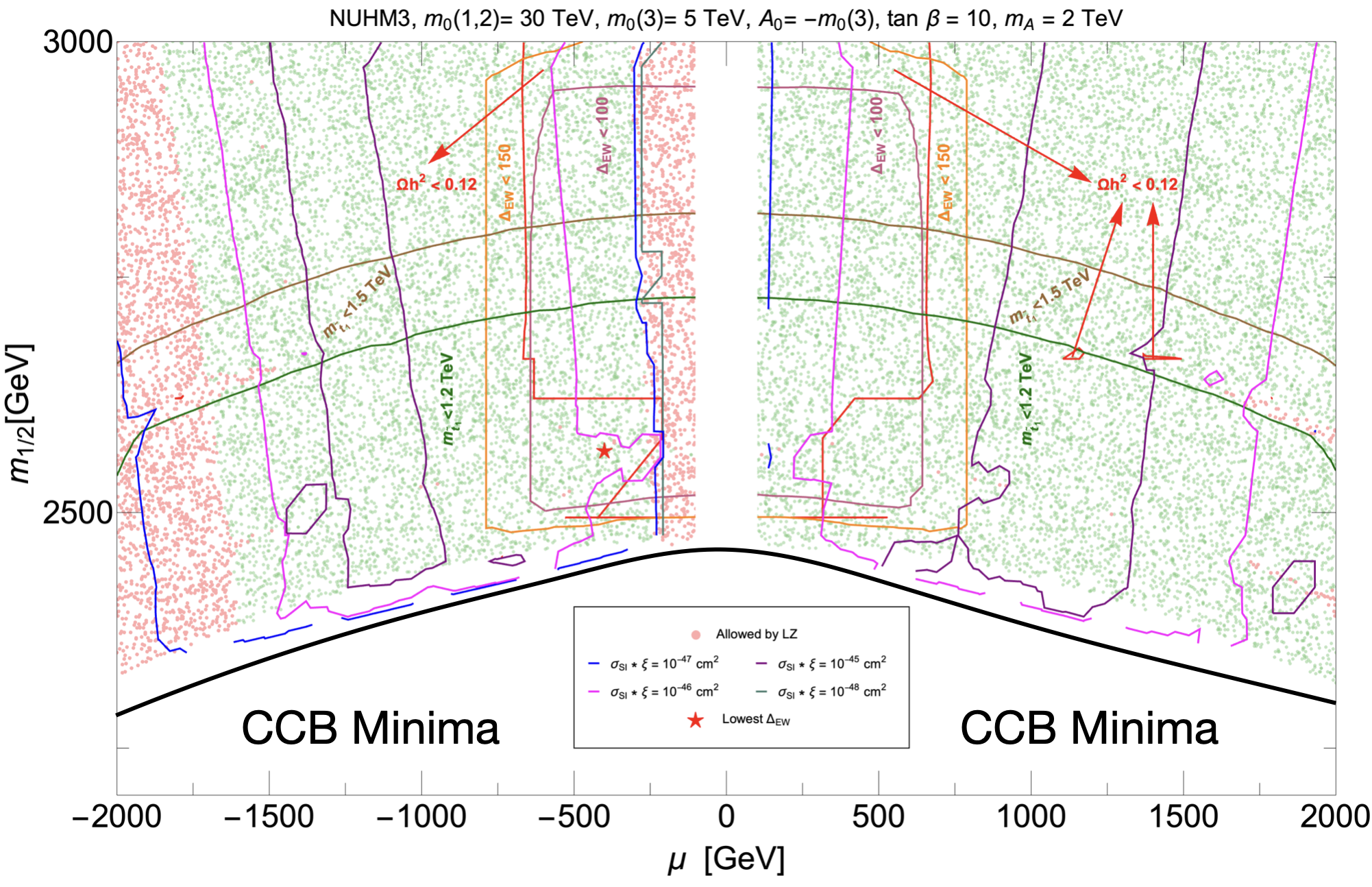}}
    \caption{The $\mu $ vs. $m_{1/2}$ parameter space of the NUHM3 model
      for $m_0=5$ TeV, $m_0(1,2)=30$ TeV,
      $A_0=-m_0(3)$, $\tan\beta =10$ and $m_A=2$ TeV.
      We show contours of $m_{\tg}$, $m_{\tst_1}$, $\Delta_{EW}$,
      $\Omega_{\tchi}^{TP}h^2$ and $\xi \sigma^{SI}(p\tchi )$.
      \label{fig:mu_v_mhf_3}}
\end{figure}

In Fig. \ref{fig:mu_v_mhf_3}, we show the $\mu$ vs. $m_{1/2}$ plane in
the NUHM3 model where the first/second generation matter scalars are
split from the third generation: $m_0(1,2)=30$ TeV so that
$m_0(1,2)\gg m_0(3)$.
This case of split generations was proposed by
Dine {\it et al.}\cite{Dine:1993np} and Cohen {\it et al.}\cite{Cohen:1996vb}
as a means to reconcile a decoupling solution to the SUSY flavor and
CP problems with naturalness which requires not-too-heavy third generation
scalars.
It is also expected to occur from generic gravity-mediation where
no symmetry exists to enforce the universality solution to the SUSY flavor/CP
problems.
It actually emerges from the strong landscape picture\cite{Baer:2019zfl}
where a power-law draw to large soft terms is balanced by the requirement
of a derived weak-scale not-too-far from our measured value
(to enforce the atomic principle
that atoms as we know them should emerge from landscape vacua in order to
generate the anthropically-required complexity for observers to
exist)\cite{Agrawal:1997gf}.

The first feature of Fig. \ref{fig:mu_v_mhf_3} that stands out is the
large unshaded region at low $m_{1/2}$ which is excluded due to the presence of
charge-and-color-breaking (CCB) minima in the scalar potential.
This is due to two-loop RG effects\cite{Baer:2024hpl} where the large
first/second generation soft terms suppress third generation soft terms,
especially $m_{U_3}^2$, to run negative at energy scale $Q\sim m_{weak}$.
This CCB region actually encompasses all the LHC-excluded $m_{\tg}\alt 2.2$ TeV
region and gives a reason why SUSY wasn't discovered at LHC Run 2 via
conventional gluino pair production reactions.
Notice also the $m_{\tst_1}<1.2$ TeV bound which cuts through the middle
of the plot, showing the large swath of parameter
space rules out by LHC top-squark searches.
The region below the $m_{\tst_1}\sim 1.5$ Tev contour should be
accessible to Run 3 and HL-LHC top-squark pair searches\cite{Baer:2023uwo},
again due to the suppression of top-squark soft terms by the 20 TeV
first/second generation scalars.

Another notable feature of Fig. \ref{fig:mu_v_mhf_3} is that LZ excludes
the vast green shaded region, but the red-shaded regions for small and
large values of $|\mu |$, where the LSP is nearly pure higgsino or gaugino,
survive due to these blind spots.
From the plot, no regions survive with $\Delta_{EW}\alt 30$.
From Ref. \cite{Baer:2024hpl}, these regions can occur,
but for smaller ranges of $m_0(3)> 5 $ TeV.
The most natural regions of this plot occur at small $|\mu |$ and lower
values of $m_{1/2}$. In these regions, there are not enough cancellations in
$\Sigma_u^u(\tst_{1/2})$ to render these terms natural even though
here $m_{\tst_1}\alt 1.2$ TeV so that the lighter top-squarks are
excluded by Run 2 LHC stop search bounds.

\section{General parameter space scan}
\label{sec:gen}

In this Section, we generalize the above NUHM2 plane to a wide variety of
parameter choices as arises from a general model scan.
We re-do the p-space scan of Ref. \cite{Baer:2018rhs}, but now containing $\mu <0$ points
as well.
\bea
m_0 &:& 0-10\ {\rm TeV},\nonumber\\
m_{1/2} &:& 0.5-3\ {\rm TeV},\nonumber\\
A_0 &:& -20 \to +20\ {\rm TeV},\\
\tan\beta &:& 4-58, \\
 |\mu | &:& 100-500\ {\rm GeV},\ \ (both\ signs) \nonumber \\
m_A &:& 0.25 - 10\ {\rm TeV}.
\eea
with $\xi =min \left[ 1,\Omega_{\tchi}^{TP}h^2/0.12 \right]$.
We restrict the scan to {\it natural} parameter space with $|\mu |<500$ GeV
since we regard unnatural models as highly disfavored.
The results of the scan are plotted in $\xi\sigma^{SI} (\tchi, p)$ vs.
$m_{\tchi}$ plane, where green points have $\mu >0$ with $\Delta_{EW}>30$ and
red points have $\mu >0$ with $\Delta_{EW}<30$ (natural).
Also, pink points have $\mu <0$ with $\Delta_{EW}>30$ while blue points
have $\mu <0$ with $\Delta_{EW} <30$.
We also display the LZ2025\cite{LZ:2024zvo} DD mass bound.
\begin{figure}[htb!]
\centering
    {\includegraphics[height=0.4\textheight]{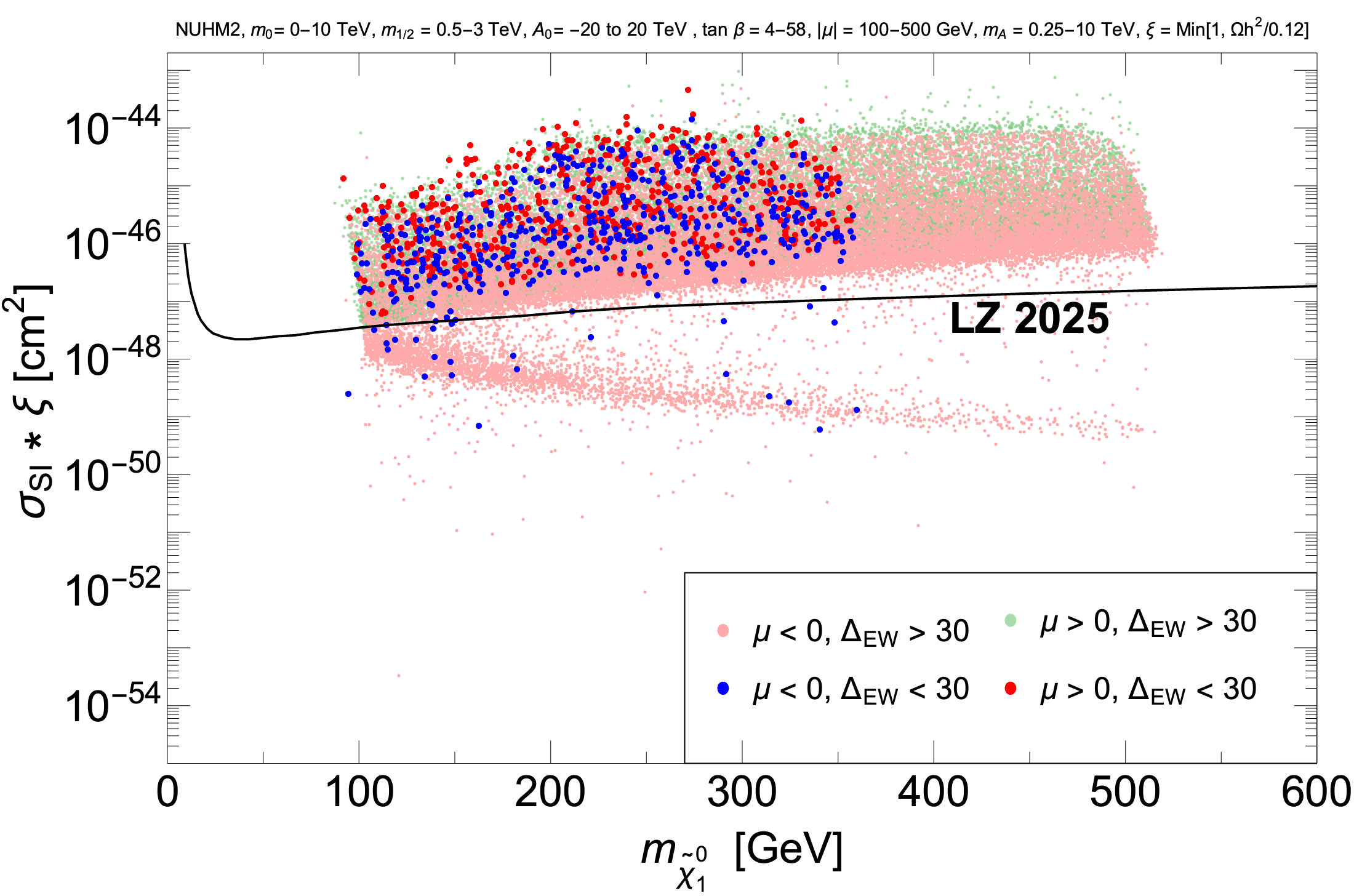}}
        {\includegraphics[height=0.4\textheight]{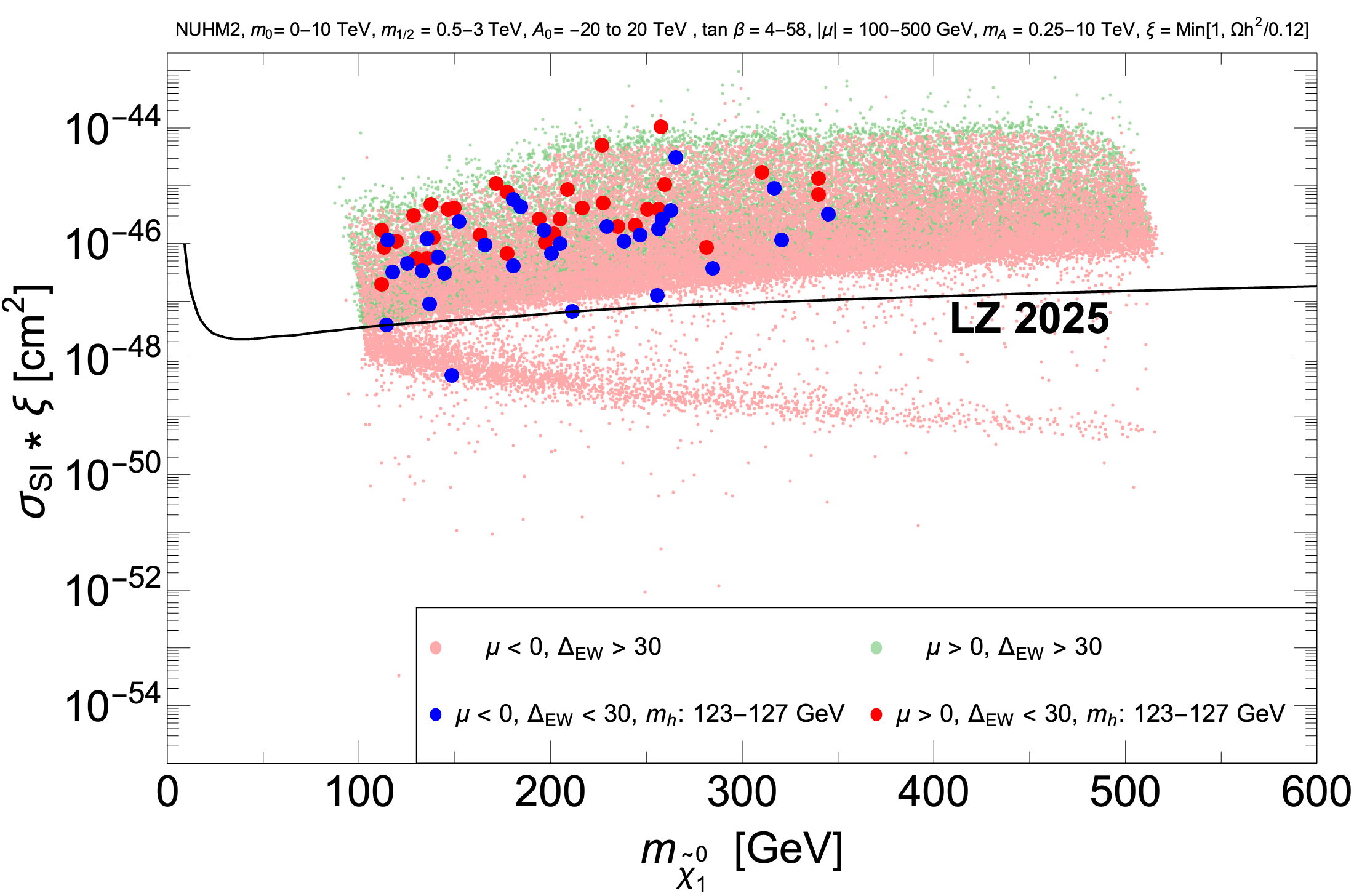}}
        \caption{Plot of {\it a}) general scan points in the
          $\xi\sigma^{SI}(\tchi p )$ vs. $m_{\tchi}$ plane, showing the LZ
          limit along with natural and unnatural SUSY scan points.
          In frame {\it b}), we show the same points but imposing
          the Higgs mass constraint $123<m_h<127$ GeV.
      \label{fig:sigSI}}
\end{figure}

From frame {\it a}),
we see that all scan points with $\mu >0$ are in fact excluded
(assuming the TP-value of $\xi$). This is expected, since with $M_{1,2}>0$ and
$\mu >0$ one has $sign (M_i/\mu )=+1$, so the bino and wino blind spot
conditions cannot be met and the SI coupling is never cancelled.
For $\mu <0$, there is a long band of
blind spot points which are not excluded by LZ but which are unnatural.
In addition, a few blue points escape the LZ TP-WIMP bounds while maintaining
naturalness.
Of these, the points with $m_{\tchi}\alt 200$ GeV are excluded by the LHC
soft dilepton search limits\cite{ATLAS:2019lng,CMS:2021edw} which extend up to $m_{\tchi_2^0}\agt 200$ GeV for
$m_{\tchi_2^0}-m_{\tchi_1^0}\sim 10$ GeV.

In frame {\it b}), we require in addition the Higgs mass constraint
$123<m_h<127$ GeV (allowing for a $\pm 2$ GeV error in the Higgs mass theory
calculation).
The previously surviving blue points with $m_{\tchi}\agt 200$ GeV
all have Higgs mass $m_h\alt 123$ GeV and so are now excluded.
This appears to follow because, in the natural region with $|\mu |\alt 350$ GeV
and $m_{1/2}\agt 500$ GeV (hence $M_1\agt 200$ GeV), the bino blind spot
$M_1=|\mu |\sin 2\beta$ cannot be reached at moderate $\tan\beta$; the only
available natural blind spot is then the higgsino case, which drives
$\tan\beta\to 1$ and so suppresses $m_h$ below its measured value.

\begin{figure}[htb!]
\centering
    {\includegraphics[height=0.5\textheight]{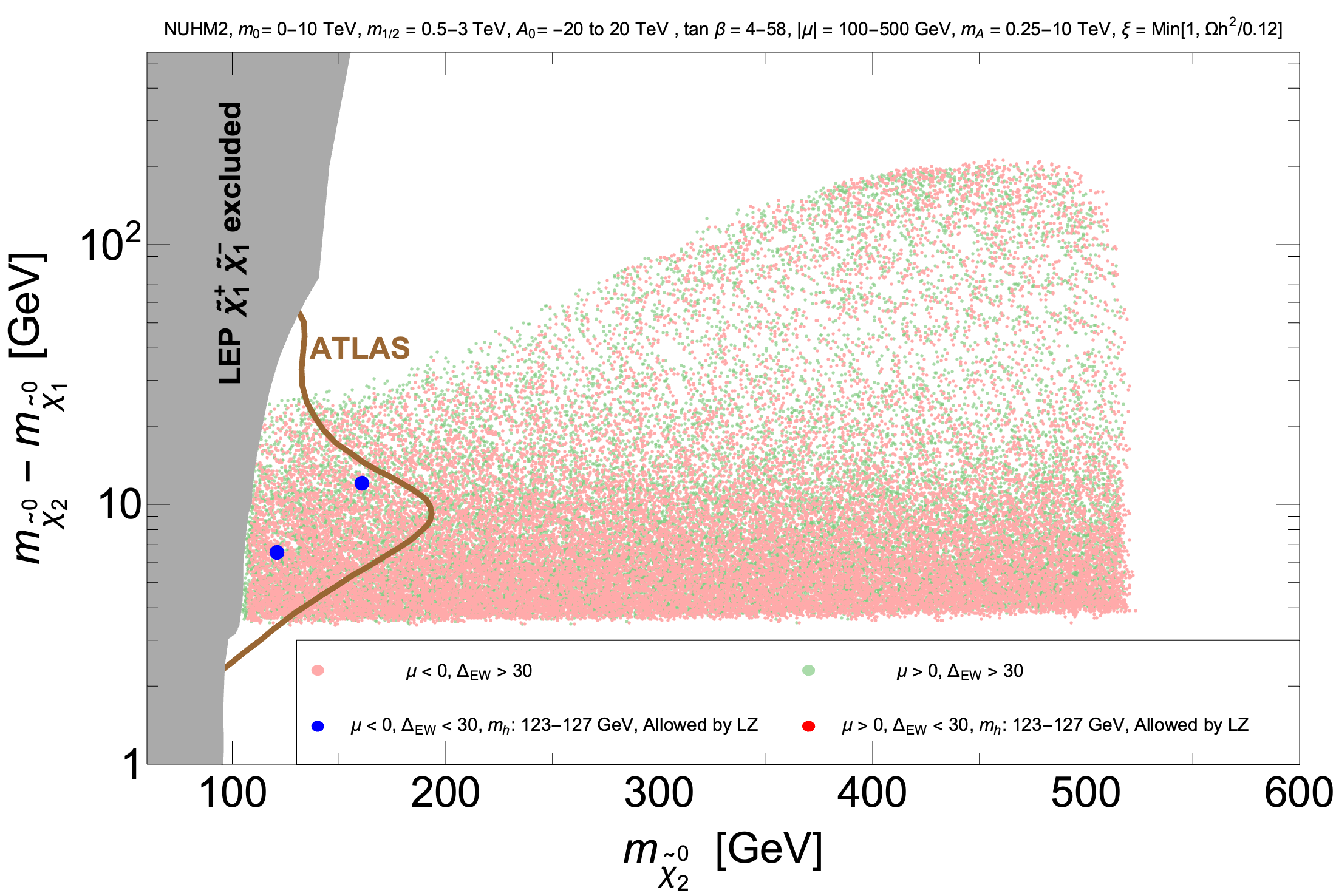}}
    \caption{Plot of general scan points in the $ m_{\tchi_2^0}-m_{\tchi_1^0}$ vs. $m_{\tchi_2^0}$ parameter space
    showing points with and without naturalness constraint ($\Delta_{EW} < $ 30), the Higgs mass constraint $123<m_h<127$ GeV
    and the LZ constraint.
      \label{fig:mz2_mz1}}
\end{figure}

In Fig. \ref{fig:mz2_mz1}, we show the general scan points in
$ m_{\tchi_2^0}-m_{\tchi_1^0}$ vs. $m_{\tchi_2^0}$ plane where the brown line depicts
the LHC soft dilepton search limits from ATLAS\cite{ATLAS:2019lng} and the
region to the left of this line is experimentally excluded.
In this plot, the green and the pink points have same meaning as in
Fig. \ref{fig:sigSI} and the red and the blue points denote those points
with  $\mu > 0$ and $\mu < 0$ respectively:
they satisfy the naturalness constraint ($\Delta_{EW} < $ 30),
the Higgs mass constraint $123<m_h<127$ GeV and the LZ constraint.
From Fig. \ref{fig:mz2_mz1}, it can be concluded that:
no points that satisfy all the three above-mentioned constraints survive
for $\mu > 0 $ and those with $\mu < 0$ that survive all the three
above-mentioned constraints are excluded by the LHC soft dilepton search
limits\cite{ATLAS:2019lng,CMS:2021edw} (as also mentioned above).

\section{Summary and conclusions}
\label{sec:conclude}

While many SUSY models suffer from the Little Hierarchy Problem, where
contributions $m_c\equiv m_Z\sqrt{\Delta_{EW}/2}$ to the weak scale are much
larger than its measured value, natural SUSY models are those whose
values of $m_c$ are comparable to $m_{weak}\sim m_{W,Z,h}\sim 100$ GeV,
and thus do not suffer from the LHP. If we require naturalness in the
QCD sector as well, then one expects the PQ solution to the strong CP problem
and then one expects mixed axion plus higgsino-like WIMP dark matter in
RPC models and perhaps all-axion-DM in RPV models.
In the former case, where WIMPs make up only a portion of the DM,
then the naive expectation is that they comprise the TP fraction $\xi$
of the total DM relic density. The $\xi$ reduction factor used to be able
to help stable higgsino-like WIMPs escape the ever-strengthening bounds from
WIMP DD experiments. It does in fact save SUSY models from IDD experiments since
these would be suppressed much more by a factor $\xi^2$\cite{Baer:2016ucr}.

However, the recent strong direct WIMP detection bounds from the LZ
experiment seem to exclude even natural SUSY models
(defined here as models with naturalness measure $\Delta_{EW}<30$)
with stable higgsino-like WIMPs that comprise just the thermally-produced
fraction of the totality of DM (the remainder being composed of SUSY-DFSZ axions with
greatly reduced $a\gamma\gamma$ coupling\cite{Bae:2017hlp}).
One way to save this scenario would be the presence of DD blind spots which
occur for gravity-mediation models with positive gaugino masses
but with $\mu <0$.
We examined this case, and find the blind spots can only save the day for
unnatural SUSY models, but not natural ones.
The natural higgsino-like stable WIMPs could still survive if there is large
entropy dumping after neutralino freeze-out but before the
onset of BBN\cite{Gelmini:2006pq,Bae:2013qr,Bae:2013hma,Bae:2014rfa,Bae:2022okh,Baer:2023bbn}.
However, this scenario seems to require rather light saxion fields $s$
or moduli fields $\phi$ so that decays to SUSY particles are suppressed or
forbidden; this case tends to make the $s$ or $\phi$ fields too long-lived
to survive BBN bounds on late decaying neutral relics.

A more appealing scenario is that the WIMPs are actually unstable.
In models with light enough axinos $\ta$ in the tens of keV-range,
one can generate the correct relic density of mixed axion/axino
dark matter\cite{Baer:2025oid}, but one must get past the SUGRA expectation that axino
mass should instead be of order the other soft terms\cite{Chun:1995hc,Kim:2012bb}.
A more motivated case emerges from solving the SUSY $\mu$ problem via discrete
$R$-symmetries and then regenerating $\mu$ via the Kim-Nilles mechanism.
In this case, one expects the appearance of higher-dimensional
RPV operators suppressed by powers of $(f_a/m_P)^n$.
For the cases with $n=1$, then the WIMPs typically decay in the early universe
before the onset of BBN, leaving all axion and no WIMP dark matter, in accord
with the strong LZ DD constraints\cite{Baer:2025oid,Baer:2025srs}.

{\it Acknowledgements:} 

V.B. gratefully acknowledges support from the U.S. Department of Energy,
Office of Science, Office of High Energy Physics,
under Award Number DE-SC0017647 and from the William F. Vilas Estate.
HB gratefully acknowledges support from the Avenir Foundation. The work of D.S. was partly supported by the European Research Council (ERC) under the European Union's Horizon 2020 research and innovation programme (Grant agreement No. 949451)


\bibliography{muneg}
\bibliographystyle{elsarticle-num}

\end{document}